\documentclass[superscriptaddress,reprint,aps,prl,amsfonts,amsmath]{revtex4-1}
\usepackage{amsmath, amssymb, amsfonts}
\usepackage{graphicx,epstopdf,bm}
\usepackage[colorlinks,bookmarks,urlcolor=blue,citecolor=blue,linkcolor=blue]{hyperref}
\usepackage[T1]{fontenc}
\usepackage{bbm} 

\newcommand{\pd}[2]{\frac{\partial #1}{\partial #2}}

\newcommand{\abs}[1]{\left\vert #1 \right\vert}

\newcommand{\explr}[1]{\exp\left[ #1 \right]}

\newcommand{\chem}[2]{ {{#1\atop\longrightarrow}\atop{\longleftarrow\atop #2}} }

\DeclareMathOperator*{\argmax}{arg\,max}
\newcommand{\Na}{\text{\tiny Na}}
\newcommand{\K}{\text{\tiny K}}
\newcommand{\rl}{\mathrm{l}}
\newcommand{\rf}{I}
\newcommand{\app}{\mathrm{app}}

\newcommand{\bx}{\mathbf{x}}
\newcommand{\bp}{\mathbf{p}}
\newcommand{\pv}{p_{v}}
\newcommand{\pw}{p_{w}}
\newcommand{\hgam}{\lambda_{M}}
\newcommand{\hgamna}{\tau_{m}}
\newcommand{\Nat}{$\text{Na}^{{+}}$}
\newcommand{\Kt}{$\text{K}^{{+}}$}

\begin{document}
\title{Breakdown of fast-slow analysis in an excitable system with channel noise}
\author{Jay M. Newby}
\email{newby.23@mbi.osu.edu}
\affiliation{Mathematical Bioscience Institute, Ohio State University, 1735 Neil Ave. Columbus, OH 43210}
\author{Paul C. Bressloff}
\affiliation{Department of Mathematics, University of Utah, 155 South 1400 East, Salt Lake City UT 84112}
\author{James P. Keener}
\affiliation{Department of Mathematics, University of Utah, 155 South 1400 East, Salt Lake City UT 84112}

\begin{abstract}
We consider a stochastic version of an excitable system based on the Morris-Lecar model of a neuron, in which the noise originates from stochastic Sodium and Potassium ion channels opening and closing. 
One can analyze neural excitability in the deterministic model by using a separation of time scales involving a fast voltage variable and a slow recovery variable, which represents the fraction of open Potassium channels. 
In the stochastic setting, spontaneous excitation is initiated by ion channel noise.
If the recovery variable is constant during initiation, the spontaneous activity rate can be calculated using Kramer's rate theory.
The validity of this assumption in the stochastic model is examined using a systematic perturbation analysis.
We find that in most physically relevant cases, this assumption breaks down, requiring an alternative to Kramers theory for excitable systems with one deterministic fixed point.
We also show that an exit time problem can be formulated in an excitable system by considering maximum likelihood trajectories of the stochastic process.
 \end{abstract}
\maketitle
Any understanding of brain function must include the role of noise.
Neural networks possess the ability to perform complex computations---taking advantage of noise when possible, while still performing reliably.
Broadly speaking, a given neuron within a network receives input from two main sources of noise: extrinsic background synaptic activity and intrinsic noise due to thermal fluctuations affecting cellular physiology.

One source of intrinsic noise is ion channel fluctuations \cite{white00a}.
Sodium (\Nat) and Potassium (\Kt) ion channels randomly shift between open and closed conformations due to the effects of thermal fluctuations, and the rate at which channels switch state depends on the membrane voltage.  
The voltage dependent activity of ion channels gives rise to membrane excitability.

Once the voltage crosses a certain threshold, a transient spike in voltage, called an action potential, is initiated.
Ion channel noise can lead to spontaneous action potentials (SAPs), which can have a large effect on network function.
If SAPs are too frequent, a neuron cannot reliably perform its computational role.
Hence, ion channel noise imposes a fundamental limit on the density of neural tissue.
Smaller neurons must function with fewer ion channels, making ion channel fluctuations more significant and more likely to cause a SAP.~
The effect of spontaneous activity on the reliability of a neuron can be quantified using information theory \cite{schneidman98a}, but the relationship between ion channel noise and spontaneous activity remains unresolved.
Ultimately, the goal is to understand the relationship between single channel dynamics, channel density, and the spontaneous activity rate.
Separately, these are experimentally accessible quantities, but channel conductances are not experimentally observable, and the dynamics of the full system must be inferred by observing the voltage.
From a theoretical perspective, the challenge is to formulate an exit time problem for a nonlinear system with only one deterministic fixed point.

Deterministic single neuron models, such as the Hodgkin-Huxley model, are useful tools for understanding membrane excitability \cite{keener09a}.
These models assume a large population of ion channels so that their effect on membrane conductance can be averaged.
As a result, the average fraction of open ion channels modulates the effective ion conductance, which in turn depends on voltage.
The Morris--Lecar (ML) model can be understood as a simplified version of the Hodgkin-Huxley model in which the fraction of open sodium channels is assumed to be in quasi-steady state so that there are two dynamical variables: the voltage $v$ and the fraction $w$ of open \Kt~channels.
The deterministic ML equation is
\begin{align}
  \label{eq:20}
  C_{\rm m}\dot{v} &= x_{\infty}(v)f_{\Na}(v) + wf_{\K}(v) + f_{\rl}(v) + I_{\app}\\
\nonumber
  \dot{w} &= (w_{\infty}(v) - w)/\tau_{w}(v),
\end{align}
where $f_{i}(v) = g_{i}(v_{i}-v)$ determine the ionic currents and $x_{\infty}(v) = (1 + \tanh(2(\gamma_{\Na}v + \kappa_{\Na})))/2$ is the fraction of open \Nat~channels
The steady state for $w$ is $w_{\infty}(v)=(1 + \tanh(2(\gamma_{\K}v + \kappa_{\K})))/2$, and the time constant $\tau_{w}(v) = 2\beta_{\K}\cosh(\gamma_{\K}v+\kappa_{\K})$ is generally assumed to be large so that the $w$ dynamics are slow compared to $v$.
We nondimensionalize voltage so that $v \to (v + v_{\rm eff})/v_{\rm{eff}}$, where $v_{\rm eff} = \frac{\abs{g_{\K}\varphi v_{\K} + g_{\rm l}v_{\rm l}}}{\abs{g_{\K}\varphi + g_{\rm l}}}$
(See Supplementary Material for parameter values.)

\begin{figure}[tbp]
  \centering
  \includegraphics[width=8cm]{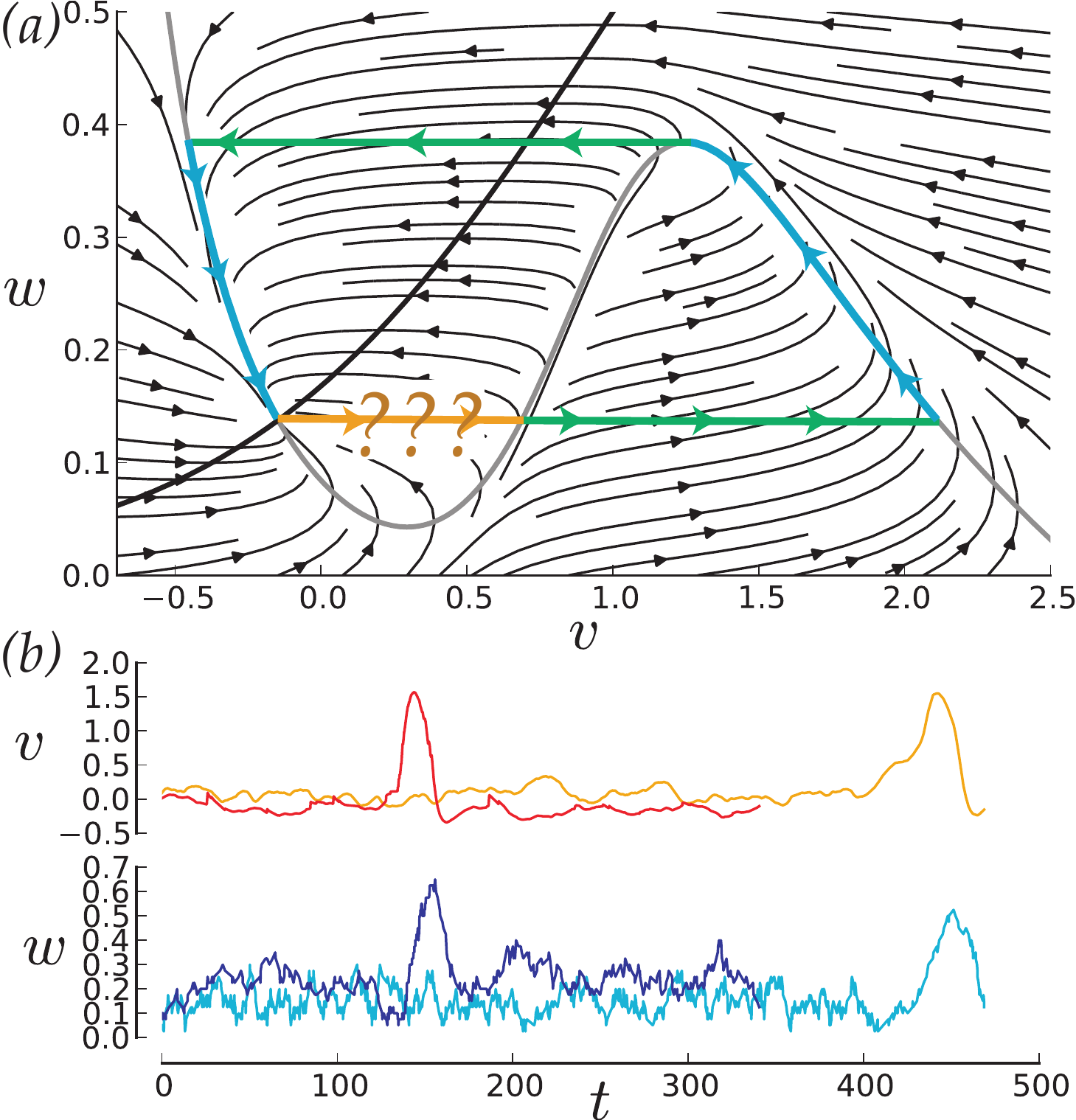}\\
  \caption{(a) Deterministic phase plane dynamics.  Nullclines: $\dot{v} = 0$ (grey) and $\dot{w}=0$ (black).  Black streamlines represent deterministic trajectories. Green/blue curves represent an action potential in the limit of slow $w$. 
The orange curve represents the central question of this letter.  What trajectory does a SAP most likely follow during initiation, and is $w$ constant on that path?
(b) Stochastic trajectories. Stochastic \Kt~and \Nat~channels: red for $v(t)$ and blue for $w(t)$.  Stochastic \Kt~and deterministic \Nat: orange for voltage $v(t)$ and light blue for $w(t)$.} 
  \label{fig:intro_a}
\end{figure}

In the ML model of a neuron \cite{morris81a}, there is no well-defined threshold for initiation of a SAP, but an effective threshold can be derived using a fast-slow analysis or separation of time scales.\cite{keener09a}.
The \Kt~channels open and close slowly compared to \Nat~channels, and the voltage response to changes in the fraction of open \Nat~channels is so fast that the fraction of open \Kt~channels, $w$, remains relatively constant.
The full system has only one fixed point (see Fig.~\ref{fig:intro_a}), but if $w$ is constant, the voltage equation is bistable with three fixed points: 
the stable resting potential, an unstable voltage threshold, and the stable excited state.
In Fig.~\ref{fig:intro_a}a, the phase plane dynamics of the system is shown along with alternating fast-slow segments of the action potential.
To initiate a SAP, noise must drive the voltage from the resting potential to a deterministic action potential trajectory.

In the stochastic setting, fixing $w$ to be constant during initiation reduces the problem to exit from a potential well, and Kramers theory provides an estimate for the spontaneous firing rate \cite{chow96a,keener11a}.
However, it is not clear if the fast-slow analysis is valid in a stochastic setting.
If $w$ is not constant then an exit time problem must be constructed for the full system, where Kramers theory does not apply.
Therefore, we are immediately faced with a dilemma if we hope to reduce the problem to an exit from a potential well.
How does one formulate an exit time problem in an excitable system with no clearly-defined threshold?
In certain limiting cases one can extrapolate a threshold called a ghost separatrix \cite{khovanov13a}, but a more broadly applicable definition has not yet been developed.

In this letter, we introduce a systematic perturbation analysis which shows that (i) $w$ is most likely not constant during initiation of a SAP and (ii) there is a well defined threshold for action potential initiation, allowing for the formulation of an exit time problem in an excitable system.

Initiation of a SAP can occur through two different mechanisms.
Assume that at the resting potential, the number of open \Nat~and \Kt~channels is set to their average value.
To study SAP initiation, one might first add \Nat~channel noise, but keep the slow \Kt~channels fixed as in the deterministic analysis.
Voltage fluctuations caused by stochastic \Nat~can drive the system from the resting potential over the threshold, initiating a SAP.
The stochastic initiation of a SAP then reduces to a familiar problem: exit from a potential well \cite{keener11a}.
Fixing $w$ constant is valid for the deterministic analysis, but even if average \Kt~channel activity is slow, how do \Kt~channel fluctuations affect SAP initiation?
Fluctuations in the number of open \Kt~channels implies the possibility that several channels close simultaneously, decreasing $w$ and thereby causing $v$ to rise.
Decreasing $w$ also reduces the voltage threshold (see Fig.~\ref{fig:intro_a}).
Indeed, Monte-Carlo simulations of the ML neuron (Fig.~\ref{fig:intro_a}b) show that SAP can be generated by \Kt~channel noise alone, without \Nat~channel noise.
Noise induced excitability has been studied in the FitzHugh--Nagumo model with white noise in the slow variable \cite{lindner99a}.
Hence, for fast/slow systems, there are two limiting cases: noise applied only to the fast variable \cite{keener11a} or only to the slow variable \cite{lindner99a}.
In this letter, noise in both variables is considered.

Past efforts to understand the relationship between SAP and ion channel noise focus on a Langevin (or diffusion) approximation.
As a first approximation, one can add white noise to a given deterministic equation, such as the ML model \eqref{eq:20}.
A better approach is to systematically derive a Langevin approximation from a more detailed model of ion channel fluctuations \cite{chow96a}.
However, as recently shown in \cite{keener11a}, Langevin approximations break down when considering metastable dynamics such as initiation of a SAP.
Moreover, both studies assume that $w$ is constant during SAP initiation.

A stochastic version of the ML model is formulated as follows.  
The voltage equation with $n = 0, 1, \cdots, N$ open \Nat~channels and $m=0, 1, \cdots, M$ open \Kt~channels is
\begin{equation}
  \label{eq:5}
  \dot{v} = \rf(v, m, n)  \equiv \frac{n}{N}f_{\Na}(v) + \frac{m}{M}f_{\K}(v) + f_{\rl}(v) + I_{\app}.
\end{equation}
We assume that each channel is either open or closed and switches between each state according to
\begin{equation}
  \label{eq:4}
  O\chem{\beta_{i}a_{i}(v)}{\beta_{i}b_{i}(v)}C,\quad i = \mathrm{Na},\;\mathrm{K},
\end{equation}
where the transition rates are $a_{\Na}(v) = e^{4(\gamma_{\Na}v + \kappa_{\Na})}$, $b_{\Na} = 1$, $a_{\K}(v) = e^{\gamma_{\K}v + \kappa_{\K}}$, and $b_{\K}(v) = e^{-\gamma_{\K}v - \kappa_{\K}}$.  
We assume that the \Nat~channels open and close rapidly, so that $1/\beta_{\Na} \ll \tau_{m}$, where $\tau_{m}=C_{\rm m}/g_{\rm L}$ is the membrane time constant.  
Taking $m$ and $n$ in \eqref{eq:5} to be stochastic birth/death processes, we obtain a stochastic hybrid process, formulated in terms of its probability density, which satisfies the Chapman--Kolmogorov equation \cite{gardiner83a},
\begin{equation}
  \label{eq:1}
  \pd{}{t}\rho(v, m, n, t) = -\pd{}{v}(\rf \rho) + \beta_{\K}\mathbb{L}_{\K}\rho + \beta_{\Na}\mathbb{L}_{\Na}\rho.
\end{equation}
The jump operators, $\mathbb{L}_{\Na} = (\mathbb{E}^{+}_{n}-1)\Omega^{+}_{\Na}(n|v) + (\mathbb{E}^{-}_{n}-1)\Omega^{-}_{\Na}(n|v)$,
and $\mathbb{L}_{\K} = (\mathbb{E}^{+}_{m}-1)\Omega^{+}_{\K}(m|v) + (\mathbb{E}^{-}_{m}-1)\Omega^{-}_{\K}(m|v)$,
govern opening/closing of \Nat~and \Kt~channels, respectively, with $\mathbb{E}^{\pm}_{a}f(a) = f(a\pm 1)$, $\Omega^{+}_{\Na}(n|v) = n$, $\Omega^{-}_{\Na}(n|v) = (N-n)a_{\Na}(v)$, $\Omega^{+}_{\K}(m|v) = m a_{\K}(v)$, and $\Omega^{-}_{\K}(m|v) =(M - m)b_{\K}(v)$.

The deterministic system \eqref{eq:20} is recovered in the limit $\beta_{\Na}\to \infty$, $M\to \infty$.
After setting $w = m/M$, the limit yields $x_{\infty}(v) = a_{\Na}(v)/(1+a_{\Na}(v))$ and $w_{\infty}(v) = a_{\K}(v)/(b_{\K}(v)+a_{\K}(v))$, which is consistent with \eqref{eq:20} \cite{keener11a}. 

A perturbation framework has been developed to study metastable activity in similar models \cite{newby11b,keener11a,newby12a}.
Similar methods have also been applied to excitable systems perturbed by white noise \cite{khovanov13a}. (For more background see \cite{ludwig75a,dykman96a,maier97a,schuss10a,tel89a}.)
The model has two large parameters, and in order to obtain a single small parameter to carry out a systematic perturbation expansion, we define $\epsilon \ll 1$ such that $\beta_{\Na}^{-1} = \tau_{m}\epsilon$ and  $M^{-1} = \hgam \epsilon$, with $\hgam = O(1)$.
Of course, $N$ could also be a large parameter, but taking the limit $N\to\infty$, $M\to\infty$ yields a different deterministic limit than \eqref{eq:20} (requiring an additional equation for the \Nat~conductance \cite{keener11a}).  We emphasize that our approximation is valid for any choice of $N>0$.

We use a WKB perturbation method to obtain a uniformly-accurate approximation of the stationary density \cite{schuss10a}, which also tells us what path a stochastic trajectory is most likely to follow during a metastable transition (i.e., a path of maximum likelihood \cite{freidlin98a,dykman96a}).
First, we assume that the stationary solution has the form
\begin{equation}
  \label{eq:14}
  \hat{\rho}(v, w, n) = r(n|v, w)\explr{-\Phi(v, w)/\epsilon},
\end{equation}
where $\Phi(v, w)$ is referred to as the {\em quasipotential} and $r(n|v, w)$ is the conditional distribution for $n$ given $v,\; w$.  
In the classic problem of exit in a double well potential, $\Phi$ is the double well potential.
More broadly, $\Phi$ is a measure of how unlikely it is for a stochastic trajectory to reach a point in phase space.
After substituting \eqref{eq:14} into \eqref{eq:1} (with $\pd{\rho}{t}=0$) and collecting terms in $\epsilon$, we find at leading order,
\begin{equation}
  \label{eq:11}
\left[\hgamna^{-1}\mathbb{L}_{\Na} + \pv + h(v, w, \pw)\right]r(n|v,w) = 0,
\end{equation}
where $\pv = \pd{\Phi}{v}$, $\pw = \pd{\Phi}{w}$, and $ h(v, w, \pw) = \frac{\beta_{\K}}{\hgam}\sum_{j=\pm}(e^{-j\hgam \pw}-1)\Omega^{\pm}_{\K}(Mw|v)/M$.

In order to solve \eqref{eq:11} for $\Phi$ and $r$, we first take $r$ to be of the form $r(n|v, w) = A^{n}/(n!(N-n)!)$.  
The constant $A$ is determined by substituting $r$ into \eqref{eq:11} to obtain a consistency expression with two terms: one linear in $n$ and one independent of $n$.  From the former we obtain $  A = a_{\Na}(v) - \frac{\hgamna}{N}(\pv g(v, w) + h(v, w, \pw))$,
where $ g(v, w) = wf_{\K}(v) + f_{\rl}(v) + I_{\app}$.
After substituting this into the remaining $n$-independent term, we obtain the nonlinear scalar PDE for $\Phi$, $ \mathcal{H}(v, w, \pd{\Phi}{v}, \pd{\Phi}{w}) = 0$, where
\begin{equation*}
  \begin{split}
    &\mathcal{H}(v, w, \pw, \pv) = (x_{\infty}f_{\Na} + g)\pv   + h(v, w, \pw) \\ 
&- \frac{\hgamna}{N}(1-x_{\infty})[(2g + f_{\Na})\pv h  + (f_{\Na} + g)g\pv^{2} + h^{2}],
  \end{split}
\end{equation*}
which can be solved using the method of characteristics \cite{schuss10a}.
Characteristics are curves $(\bx(t), \bp(t))$ (with $\bx = (v, w)$ and $\bp = (\pv, \pw) = \nabla_{\bx}\Phi$) that satisfy the following dynamical system,
\begin{equation}
  \label{eq:19}
  \dot{\bx} = \nabla_{\bp}\mathcal{H}(\bx, \bp), \quad \dot{\bp} = -\nabla_{\bx}\mathcal{H}(\bx, \bp).
\end{equation}
Note that the deterministic system \eqref{eq:20} is recovered by setting $\bp = 0$.
Characteristic projections, $\bx(t)$, referred to as {\em metastable trajectories}, are paths of maximum likelihood leading away from the fixed point \cite{freidlin98a,dykman96a}.
The {\em action}, $\Phi(t)$, satisfying $\dot{\Phi}(t) = \mathbf{p}(t)\cdot \dot{\bx}(t)$, is a strictly increasing function of $t$, and the quasipotential is given by $\Phi(v, w) = \Phi(t)$ at the point $(v, w) = \mathbf{x}(t)$.
Note that $\dot{\Phi}=0$ along deterministic trajectories.
We solve \eqref{eq:19} using numerical ODE integration \cite{newby12a}.
A Comparison of the WKB approximation to Monte-Carlo simulations can be found in Supplementary Material.

\begin{figure}[tbp]
  \centering
  \includegraphics[width=8cm]{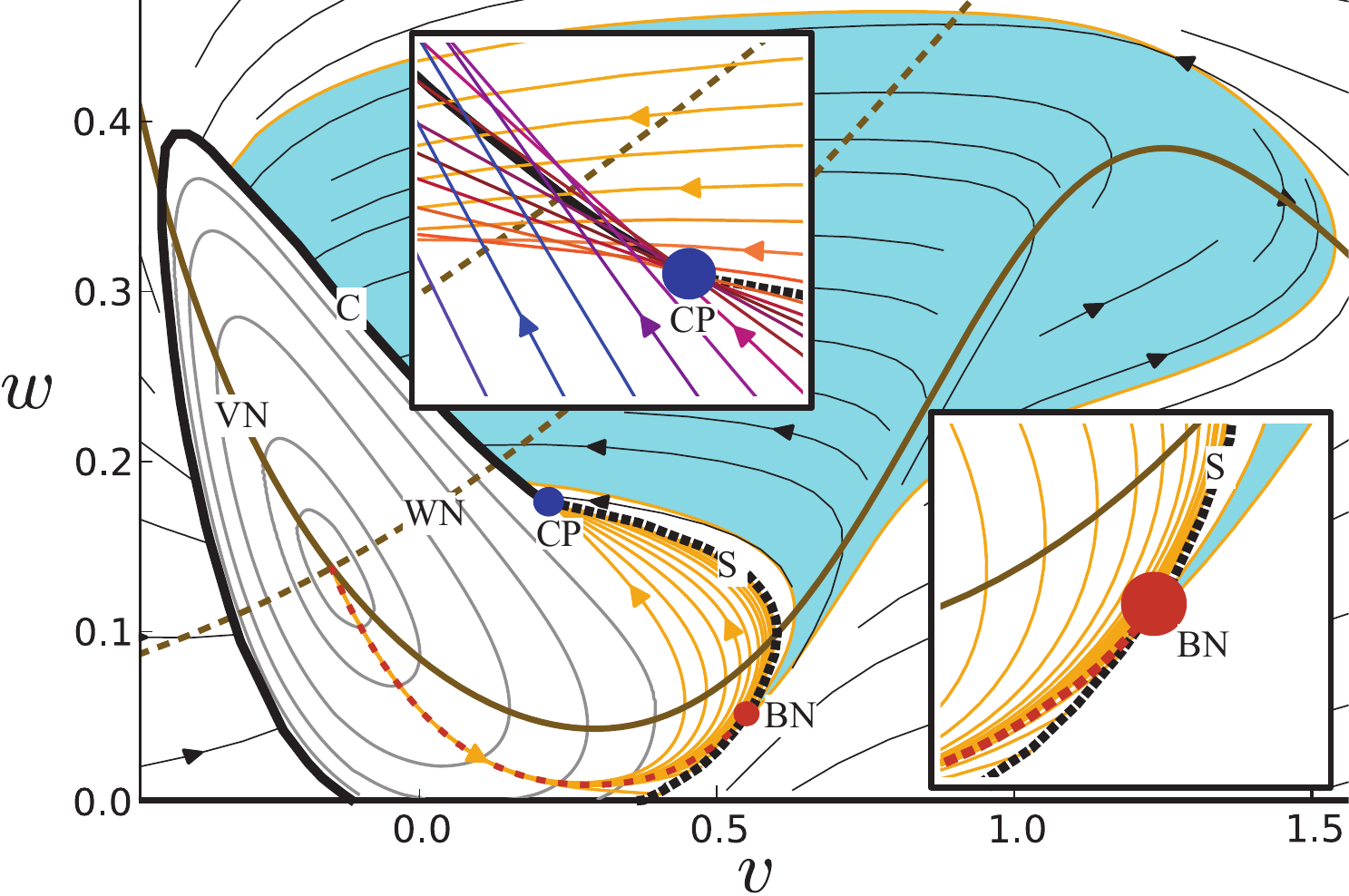}
  \caption{Orange curves are SAP trajectories, shown until they reach the metastable separatrix (S).
The dashed red curve is a SAP that reaches S near the bottleneck (BN).
All of the SAP trajectories that enter the shaded region are visually indistinguishable from the dashed red line before crossing S. 
Deterministic trajectories are shown as black streamlines.
Left inset: close up of the caustic formation point (CP) with overlapping metastable trajectories.
Level curves of $\Phi$ are shown inside the potential well region with grey lines.
Also shown are the caustic (C), $v$ nullcline (VN), and $w$ nullcline (WN).  Parameter values are $N=M=40$ and $\hgam=0.25$.}
  \label{fig:actionb}
\end{figure}
Surrounding the stable fixed point, $\Phi$ takes the shape of a potential well (Fig.~\ref{fig:actionb}), with convex level curves (grey lines).
Once $\Phi$ reaches a threshold, a caustic is formed as the solution surface folds over on itself.
Metastable trajectories begin to overlap, and the solution $\Phi(v, w)$ loses uniqueness (left inset Fig.~\ref{fig:actionb}).
Within this region, uniqueness is achieved at each point by minimizing the action over all metastable trajectories that pass through that point.
The caustic is a line along which every point is connected to two equally likely metastable trajectories; it forms an incomplete boundary around part of the potential well.
The remaining boundary is the curve of constant $\Phi(v, w) = \Phi_{c}$, where $\Phi_{c}\approx 1.034$ is the quasipotential at the caustic formation point (dashed line Fig.~\ref{fig:actionb}).
We refer to this curve as the {\em metastable separatrix}.

We identify SAP trajectories as those metastable trajectories that cross the separatrix.
SAP trajectories begin at the fixed point as a single trajectory and then fan out just before reaching the metastable separatrix (Fig.~\ref{fig:actionb}).
After crossing the separatrix, all of the SAP trajectories eventually reach the caustic.
Although all SAPs are equally likely to reach the separatrix, their likelihood of reaching the caustic depends on their amplitude.
Large amplitude SAPs are less likely and reach the caustic far from the caustic formation point.
Strictly speaking, the most probable SAP strikes the caustic formation point, but $\Phi$ increases by a very small amount in the shaded region of Fig.~\ref{fig:actionb} because SAP trajectories are very close to deterministic trajectories (black streamlines).
(The relative difference is $\abs{\Delta\Phi}/\Phi_{c}\approx 0.01$.)
Hence, the stationary density \eqref{eq:14} is nearly constant in the shaded region.

SAPs that cover the shaded region cross a very small segment of the separatrix, the center of which acts as a bottleneck for SAPs.
The shaded region represents the most likely, experimentally observable SAP trajectories; it excludes small amplitude SAPs that (crossing above the bottleneck) strike very close to the caustic formation point and the less probable SAPs that (crossing below the bottleneck) strike the caustic above or behind the potential well region.
The portion of the SAP trajectory between the fixed point and the bottleneck (see Fig.~\ref{fig:actionb} dashed curve) represents the initiation phase; it is not constant and remains below the $v$ nullcline. 

\begin{figure}[tbp]
  \centering
  \includegraphics[width = 8cm]{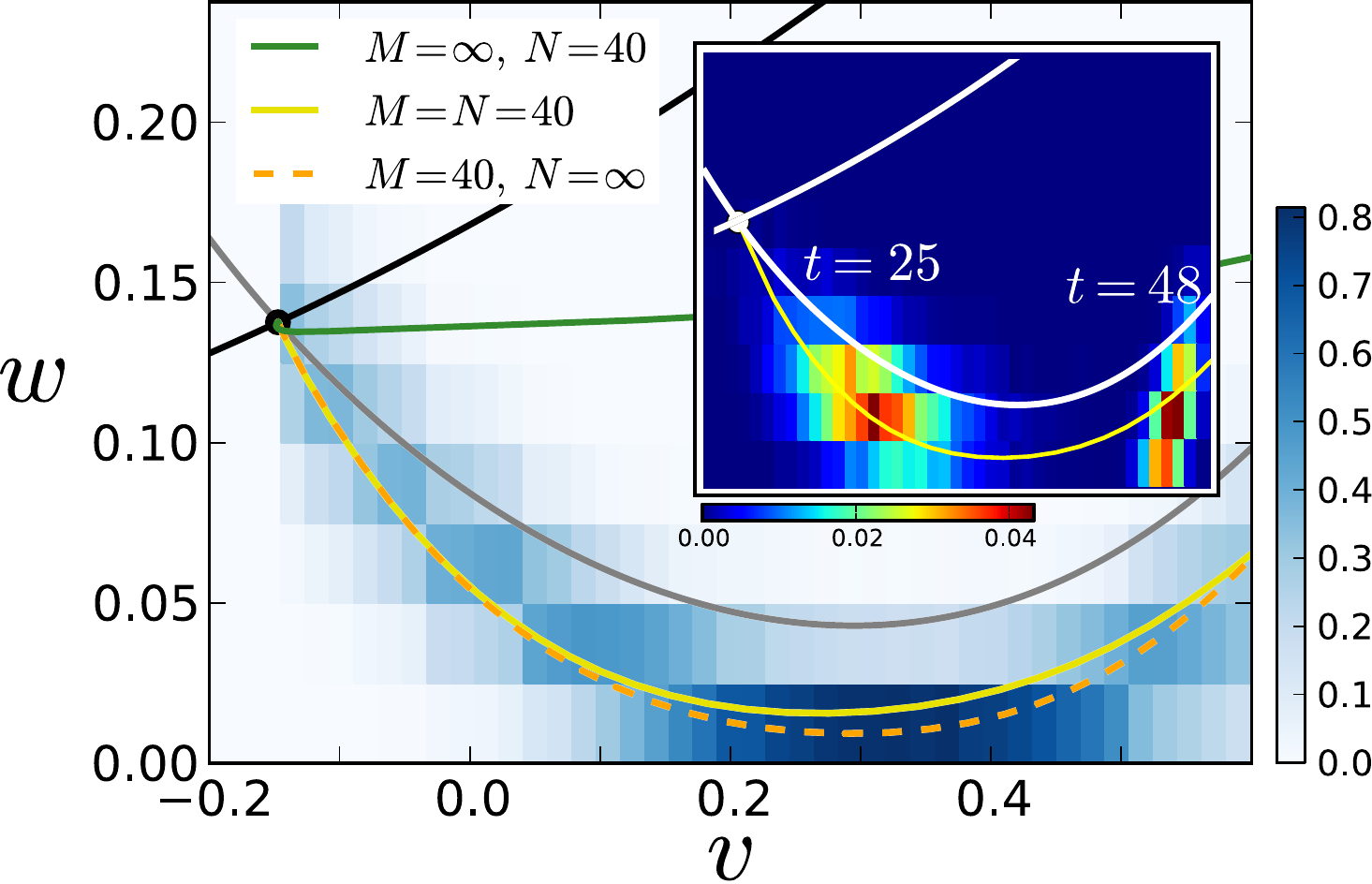}
  \caption{SAPs that pass through the bottleneck. Monte-Carlo simulations (see supplementary material) are used to obtain histograms of the path history \cite{dykman96a}, $Q(v, w, t)\frac{dv}{M} \equiv \text{Pr}[v(t) \in (v, v+dv), \,w(t) = w \,|\, v(50)=0.6,\,v(-\infty)=v_{0},\, w(-\infty)=w_{0}]$, $0<t<50$, for $M=N=40$. The heat map shows the conditional probability, $\text{Pr}[w(t)=w\, |\, v = v_m(t)]$, where $(v_{m}, w_{m}) = \argmax Q$. The argmax of each histogram is a point on the maximum likelihood trajectory and agrees with the theory (yellow line). Inset: snapshots (heat maps) of $Q$ at $t=25$ and $t=48$.}
  \label{fig:3}
\end{figure}
We have two limiting cases (see Fig.~\ref{fig:3}): (a) $N\to \infty$ and (b) $M\to \infty$.
In both cases $\mathcal{H}$ is simplified, but \eqref{eq:19} remains four dimensional.
Suppose we also assume $\beta_{\K} \ll \tau_{m}^{-1}$.
Then, for case (a), we set $x_{\infty}(\hat{v})f_{\Na}(\hat{v}) = - g(w, \hat{v}),\; \pv = 0$ and $\mathcal{H}$ reduces to $h(\hat{v}(w), w, \pw) = 0$, for which $\pw = \hgam^{-1}\ln\frac{\Omega_{\K}^{+}}{\Omega_{\K}^{-}}$ is a solution.
This solution generalizes the result in \cite{lindner99a} to channel noise.
For case (b), we recover the problem solved in \cite{keener11a}.
Fig.~\ref{fig:3} shows the SAP during the initial phase for case (a), (b), and $N = M = 40$.

To summarize our results, we find that fluctuations in the slow recovery dynamics of \Kt~channels significantly affect spontaneous activity in the ML model.
The maximum likelihood trajectory during initiation of a SAP can be thought of as a path of least resistance, dropping below the voltage nullcline where voltage increases deterministically.
Hence, SAP initiation is more likely to occur using the second of the two mechanisms mentioned in the introduction: a burst of simultaneously-closing \Kt~channels causes $v$ to increase.
If one takes $w$ to be constant, only the first mechanism is available and the path is artificially constrained, which alters the quasipotential.
In other words, constraining the path alters the effective energy barrier for SAP initiation, which significantly affects determination of the spontaneous firing rate.
Although it is more difficult to construct an exit time problem in an excitable system, this can now be done using the metastable separatrix. 
The methods used here are general and may lead to future studies of noise-induced dynamics in other nonlinear stochastic systems.
In particular, it would be interesting to extend the current analysis to the Hodgkin-Huxley model, where the \Nat~channels have a slow inactivating component.

\renewcommand{\theequation}{S.\arabic{equation}}
\renewcommand{\thefigure}{S.\arabic{figure}}
\section{Supplementary: Monte-Carlo simulations}

\begin{figure*}[tbp]
  \centering
  \includegraphics[width = 18cm]{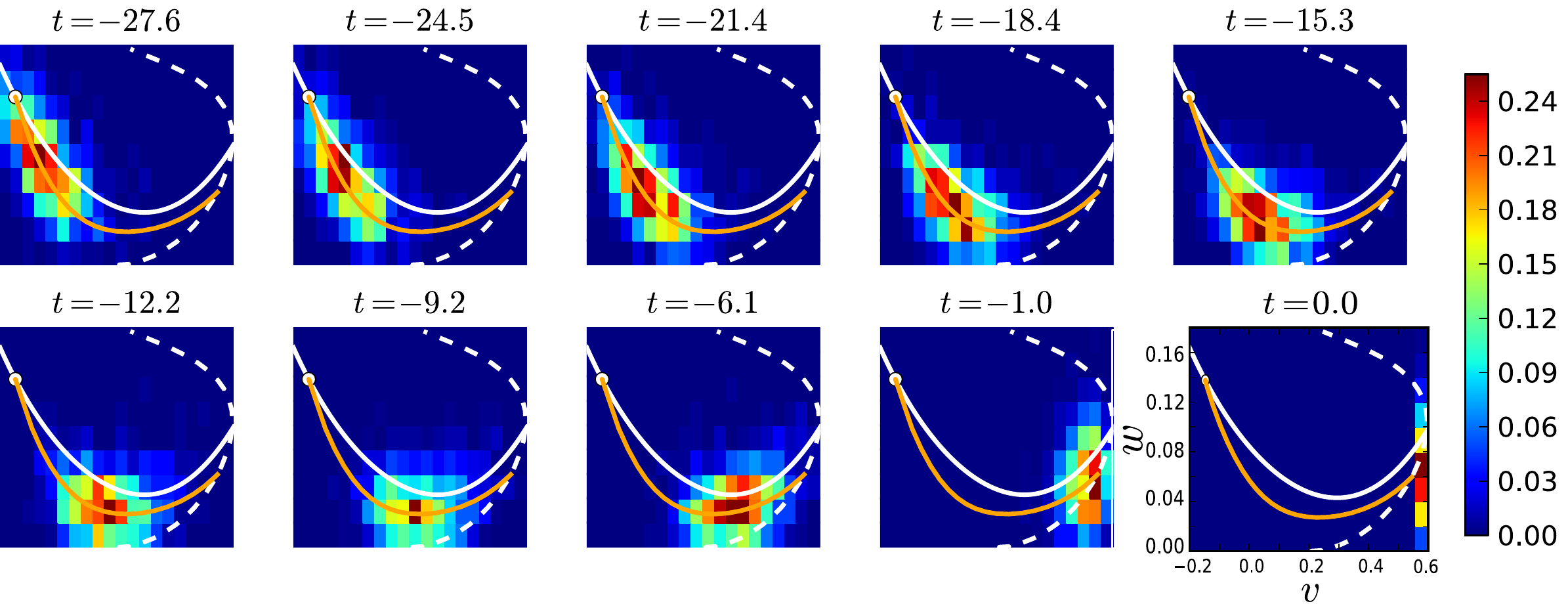}
  \caption{Histogram of Monte-Carlo trajectories prior to reaching the line $v = 0.6$ (the right edge of each pane).  Each pane shows the histogram at a different time, with $t=0$ the time at which the trajectory reaches $v = 0.6$. The orange curve shows a characteristic projection that passes through the bottleneck.  The dashed white line is the metastable separatrix.  The white dot is the fixed point.  The solid white line is the $v$-nullcline.  Parameter values used are $N = 4$, $M=50$, and $\hgam = 1/5$.  The histogram is divided into bins $(w_{m}, v_{j}) = (\frac{m}{M}, -0.2 + j\frac{0.8}{20})$, $m=0, 1, \cdots , M$ and $j = 0, 1, \cdots, 20$.}
  \label{fig:mc}
\end{figure*}

In this supplementary material, we discuss verification of our analytical/numerical results by comparison to Monte-Carlo simulations. 
Parameter values used here and in the main article are listed in \footnote{$v_{\Na} = 120 {\rm mV}$, $g_{\Na}= 4.4{\rm mS/cm}^2$, $v_{\mathrm{K}}= -84 {\rm mV}$, $g_{\mathrm{K}}= 8 {\mathrm{mS}/\mathrm{cm}}^2$, $v_{\rl}= -60{\rm mV}$, $g_{\rl}= 2{\mathrm{mS}/\mathrm{cm}}^2$, $C_{\rm m} = 20m{\rm F/cm}^2$, $\beta_{\K} = 0.02 {\rm ms}^{-1}$, $I_{\rm app}= 0.06C_{\rm m}v_{\rm eff}$, $\varphi = -0.1$, $v_{\rm eff} = 52.8 {\rm mV}$, 
$\gamma_{\Na} = 1.22/v_{\rm eff}$, $\kappa_{\Na}= -1.188 + 1.22 v_{\rm eff}$, $\gamma_{\K}= 0.8/v_{\rm eff}$, $\kappa_{\K}= 0.8 + 0.8 v_{\rm eff}$, $\tau_{m} = 10 {\rm ms}$.
}.
Motivated by Ref.~\cite{dykman96a}, we perform Monte-Carlo simulations (for details about the algorithm, see the next section) to obtain trajectories that start at the fixed point and eventually reach the line $v = 0.6$ (the right edge of each pane in Fig.~\ref{fig:mc}).
From the ensamble of these trajectories, we determine the statistics of the position of trajectories as a function of time preceding arrival at $v = 0.6$.
We then set the time at which each trajectory ends (i.e., the time at which they reach $v = 0.6$) to $t=0$ and look backward in time in order to observe the behavior of the process during the initiation phase of a spontaneous action potential.  
Trajectories are sampled to obtain histograms of the path history defined as \\

\begin{widetext}
\begin{equation}
  \label{eq:8}
  Q(v, w, t)dv/M \equiv \text{Pr}[v(t) \in (v, v + dv), w(t) = w \,|\, v(t_{f})=v_{f},v(t_{0})=v_{0}, w(t_{0})=w_{0}],\quad t_{0}<t<t_{f},\; v < v_{f}.
\end{equation}
\end{widetext}
Each pane in Fig.~\ref{fig:mc} is the histogram of $Q$ from $1\times 10^{3}$ trials at different points in time, with $v_{f}=0.6$, $t_{f} = 0$, and (effectively) $t_{0} = - \infty$ (trajectories take a long time to reach $v_{f}$).  For example, the first pane is the histogram of trajectories at $t = -27.6 {\rm ms}$ before reaching $v = 0.6$.  
The maximum likelihood trajectory is by definition the peak of the histogram as a function of time, and it is evident from Fig.~\ref{fig:mc} that it coincides with the characteristic projection (shown in orange).

Fig.~3 in the main text shows a heat map of the histogram of the conditional probability
\begin{equation}
  \label{eq:11}
  \text{Pr}[w(t)=w |v = v_m(t)] = \frac{Q(w, v_{m}, t)}{\sum_{m=0}^{M}Q(m/M, v_{m}, t)},
\end{equation}
which shows the distribution of $w$ conditioned on $v = v_{m}(t) \equiv \argmax_{v} Q(w, v, t)$.
Hence, the mode (arg max of the histogram) of the conditional probability shown in Fig.~2 for each fixed value of $v$ corresponds to the time dependent mode of the probability density $Q$ show in Fig.~\ref{fig:mc}.

\begin{figure*}[tbp]
  \centering
  \includegraphics[width=18cm]{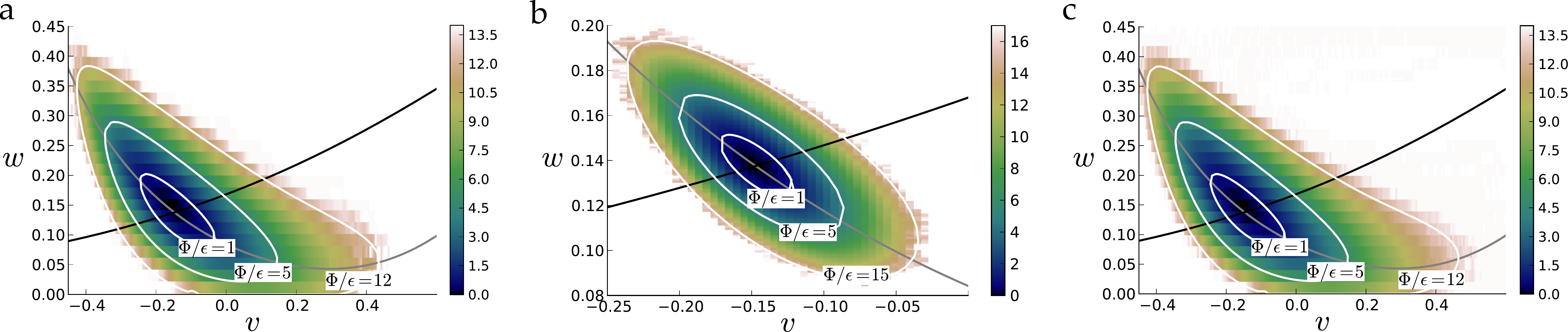}
  \caption{Comparison of the WKB approximation of the stationary density, $\hat{\rho}(v, w)$, to Monte-Carlo simulations.  The stationary density is represented as $-\log(\hat{\rho})$.  The WKB method approximates this quantity with $\Phi(v, w)/\epsilon \sim - \log(\hat{\rho}(v, w))$.  Level curves of $\Phi$ are obtained by integration of (7) and are shown (white lines) along with heat maps of histograms generated from Monte-Carlo simulations. Three cases are shown: (a) $N=1000$, $M=50$, (b) $N=50$, $M=1000$, and (c) $N = 50$, $M = 50$. We set $\epsilon = 0.1$, other parameters can be found in [1]. The histograms are divided into bins $(w_{m}, v_{j}) = (\frac{m}{M}, -0.6 + j\frac{2.6}{500})$, $m=0, 1, \cdots , M$ and $j = 0, 1, \cdots, 500$.  The $v$-nullcline (grey curve) and the $w$-nullcline (black curve) are also shown}
  \label{fig:mcstat}
\end{figure*}

The WKB method presented in the article provides an approximation of the stationary probability density function (5).  
In Fig.~\ref{fig:mcstat}, we compare this approximation to histograms obtained by Monte-Carlo simulations for three different limiting cases: (a) $N\to \infty$, (b) $M\to \infty$, and (c) $M = N = 50$.
Hence, the quasipotential from the WKB approximation is related to the stationary probability density by $\Phi(v, w) \sim -\epsilon \log(\hat{\rho}(v, w))$.
Level curves of $\Phi$ (white curves) from the WKB approximation show that $\Phi/\epsilon$ and $-\log(\hat{\rho})$ (shown as heat maps) are in close agrement.

\section{Supplementary: Monte-Carlo simulation algorithm}
Monte-Carlo simulations are generated using an extension of the algorithm presented in \cite{keener11a}.  Instead of using the Gillespie algorithm as in \cite{keener11a}, we use the next reaction method along the lines of \cite{bokes13a}.
The algorithm is exact in the sense that the transition times can be approximated to any desired precision.
The simulations were coded in C (using the GNU Scientific Library for random number generators) and carried out in Python, using the SciPy package.
In between each jump in the number of open channels, the voltage is evolved according to the deterministic dynamics 
\begin{equation}
  \label{eq:4}
  \dot{v} = \frac{n}{N}f_{\Na}(v) + \frac{m}{M}f_{\K}(v) + f_{\rl}(v) + I_{\app},
\end{equation}
The solution provides the relationship between voltage and time,
\begin{equation}
  \label{eq:100}
  v(t) = \left(v(t_{0}) - \frac{c_{2}}{c_{1}}\right)e^{-c_{1}(t-t_{0})} + \frac{c_{2}}{c_{1}},
\end{equation}
where
\begin{align}
  \label{eq:5}
  c_{1} &= \frac{n}{N}g_{\Na} + \frac{m}{M}g_{\K} + g_{\rl}, \\
  c_{2} &= \frac{n}{N}g_{\Na}v_{\Na} + \frac{m}{M}g_{\K}v_{\K} + g_{\rl}v_{\rl} + I_{\app}.
\end{align}
To compute the next jump time, we compute four random jump times for each of the four possible transitions: $n\to n\pm 1$ and $m \to m \pm 1$. 
Each transition time is distributed according to
\begin{align}
  \label{eq:1}
  W^{-}_{\Na}(t) &=1 - e^{-\beta_{\Na}n (t - t_{0})}, \\
  W^{+}_{\Na}(t) &= 1- \explr{-\beta_{\Na}\int_{t_{0}}^{t}\Omega^{+}_{\Na}(v(\tau))d\tau},\\
  W^{\pm}_{\K}(t) &= 1- \explr{-\beta_{\K}\int_{t_{0}}^{t}\Omega^{\pm}_{\K}(v(\tau))d\tau}.
\end{align}
After integrating the voltage dependent transition rates we obtain for $(i=+, j = {\rm Na})$ and $(i=\pm, j = {\rm K})$,
\begin{equation}
  \label{eq:2}
  \int_{t_{0}}^{t}\Omega^{i}_{j}(v(\tau))d\tau = \frac{1}{c_{1}}\Omega^{i}_{j}(\frac{c_{2}}{c_{1}})(E_{\rm i}(z^{i}_{j}e^{-c_{1}(t-t_{0})}) - E_{\rm i}(z^{i}_{j})),
\end{equation}
where
\begin{align}
  \label{eq:6}
  z^{+}_{\Na} &=  4\gamma_{\Na}\left(v(t_{0}) - \frac{c_{2}}{c_{1}}\right), \\
  z^{\pm}_{\K} &= \pm \gamma_{\K}\left(v(t_{0}) - \frac{c_{2}}{c_{1}}\right), 
\end{align}
and $E_{\rm i}$ is the exponential integral function defined as the Cauchy principle value integral,
\begin{equation}
  \label{eq:3}
    E_{\rm i}(x) = \int_{-\infty}^{x}t^{-1}e^{t}dt, \quad x\neq 0.
\end{equation}
Denote the jump times by $t^{i}_{j}$, $i = \pm$ and $j = {\rm Na}, {\rm K}$, and let $U$ be a uniform random variable.
The jump times are given by the solution to 
\begin{equation}
  \label{eq:10}
  W^{i}_{j}(t^{i}_{j}) = U.
\end{equation}
There is one voltage independent jump time,
\begin{equation}
  \label{eq:7}
  t^{-}_{\Na} = -\frac{\log(U)}{n\beta_{\Na}}.  
\end{equation}
Because three of the transition rates depend on voltage, and therefore time, the distributions for the jump times are not explicitly invertible.
Hence, the next jump times are given implicitly by
\begin{gather}
\nonumber
\frac{1}{c_{1}}\Omega^{+}_{\Na}(\frac{c_{2}}{c_{1}})(E_{\rm i}(z^{+}_{\Na}e^{-c_{1}(t^{+}_{\Na}-t_{0})}) - E_{\rm i}(z^{+}_{\Na})) = -\log(U), \\
  \label{eq:101}
\frac{1}{c_{1}}\Omega^{\pm}_{\K}(\frac{c_{2}}{c_{1}})(E_{\rm i}(z^{\pm}_{\K}e^{-c_{1}(t^{\pm}_{\K}-t_{0})}) - E_{\rm i}(z^{\pm}_{\K})) = -\log(U).
\end{gather}
To generate the voltage dependent jump times, a Newton root finding algorithm is applied to \eqref{eq:101} with a tolerance of $ 10^{-8}$.  
Once all four transition times have been computed, the next transition time is 
\begin{equation}
  \label{eq:9}
  t^{i_{*}}_{j_{*}} = \min_{i=\pm,j=\Na,\K}\{t_{j}^{i}\}.
\end{equation}
The global time is updated with $t \leftarrow t + t^{i_{*}}_{j_{*}}$.  The state is updated with $v \leftarrow v(t^{i_{*}}_{j_{*}})$ (where $v(t)$ is given by \eqref{eq:100} with $t_{0}$ the time of the previous jump), $n \leftarrow n + i_{*}$ if $j_{*} = {\rm Na}$, and $m \leftarrow m + i_{*}$ if $j_{*} = {\rm K}$.
\vfill
\bibliography{lib}
\end{document}